\documentstyle[12pt]{article}
\newcommand{\roughly}[1]{\raise.3ex\hbox{$#1$\kern-.75em
\lower1ex\hbox{$\sim$}}}

\begin{document}

\centerline{\Large\bf Black holes in the lab?}
\bigskip
\bigskip
\centerline{{\bf Steven B. Giddings}\footnote{email address:  
giddings@physics.ucsb.edu.  
On leave from 
Department of Physics, University of
California, Santa Barbara, CA\ 93106}}
\bigskip
\bigskip
\smallskip
\centerline{Department of Physics and SLAC,
Stanford University, Stanford,
CA\ 94305/94309}
\bigskip

\begin{abstract}

If TeV-scale gravity describes nature, black holes will be produced in
particle accelerators, perhaps even with impressive rates at the Large
Hadron Collider.  Their decays, largely via the Hawking process, will be
spectacular.  Black holes also would be produced in cosmic ray collisions
with our atmosphere, and their showers may be observable.  Such a scenario
means the end of our quest to understand the world at shorter distances, but
may represent the beginning of the exploration of extra dimensions.

\end{abstract}
\newpage

Black holes are perhaps the most profoundly mysterious objects in physics.  
We have long pondered what happens at the core of a
black hole.  The answer likely involves radically new physics, including
breakdown of space and time, and is still beyond the reach of
current approaches to quantum gravity such as string theory.  
Moreover, 
Hawking's 
discovery of black hole radiance\cite{Hawking:sw} 
and proposal that black holes violate
quantum mechanics\cite{Hawking:ra} has led us to the sharp paradox of
black hole information, which drives at the 
very heart of the problem of reconciling
quantum mechanics and gravity.  There is no clear way out:\footnote{For
reviews, see 
\cite{Giddings:1995gd,Strominger:1994tn,Giddings:1994pj,tHooft:ij}.} 
information loss
associated with breakdown of quantum mechanics apparently leads
to disastrous violations of energy conservation; information cannot escape
a black hole without violating locality; and the third alternative,
black hole remnants, lead to catastrophic instabilities.  String theorists
have recently investigated the second alternative, via holography,
but the jury is still out as no one has
managed to understand how holographic theories can reproduce the
approximately local physics that we see in our everyday world.

Experimental clues to the physics of black hole decay would be 
welcome.  Unfortunately, manufacture of microscopic black holes apparently
requires
scattering energies above the four-dimensional 
Planck mass, $M_4\sim 10^{19}$ GeV,
placing this possibility far in our future.

However, recently there has been a revolution in thinking about the
relationship between the Planck scale and the weak scale, $M_W\sim 1$TeV.
The longstanding ``hierarchy problem'' is to explain the large ratio of
these; one would naturally expect $M_W\sim M_4$.  The new idea is that the
weak scale and the {\it fundamental} Planck scale, $M_P$, are indeed the
same size, but four-dimensional gravity is weak (hence $M_4$ is large) due to 
dilution of gravity in large or warped extra dimensions.  

Specifically, for a general Poincar\'e-invariant metric 
\begin{equation}
ds^2 = e^{2A(y)} dx^{\mu2} + g_{mn}(y) dy^m dy^n\ ,\ 
\end{equation}
where $x^\mu$ are the four dimensions we see, $y^m$ are the extra
$n$ dimensions, and the function $A(y)$ is the {\it warp factor},
$M_4$ is given by
\begin{equation} 
\frac{M_4^2}{M_P^2} = M_P^n \int d^ny \sqrt{g} e^{2A}\ .
\end{equation}
Either large volume or large $A$ produce a big ratio
between $M_P$ and the observed Planck scale.  
The hierarchy problem morphs into that of explaining
why the extra dimensions are large, or highly warped.  For $n=2$--$6$ their
size ranges from $mm-fm$.  This presents a conflict with
precision measurements of the gauge forces which have now reached the scale
of $10^{-3}fm$, but we are saved by the brane world idea, which follows
naturally 
from string theory:  gauge
forces, corresponding to open strings, propagate on a brane within the 
extra dimensions, whereas gravity, which is always
transmitted by closed strings, propagates in all of the dimensions.  Such
scenarios go by the name ``TeV-scale gravity.''

If TeV-scale gravity describes nature, the consequences are astounding.  We
will begin to explore quantum gravity, and possibly string theory, at
accelerators in the relatively near future.  Indeed, model independent
bounds based on present experiments merely indicate $M_P\roughly> 800$GeV;
TeV-scale gravity could be the physics of the Large Hadron Collider (LHC).  

The most generic and spectacular result of such a scenario would be
the production of black holes in particle
accelerators\cite{Giddings:2001bu,Dimopoulos:2001hw}. 
LHC will collide
protons, which are aggregates of partons (quarks and gluons).  If 
$M_P\sim{\cal O}(TeV)$, then parton collisions with
significantly higher center-of-mass energy $E$ should produce black holes; 
we consider $E\roughly>5M_P$, such that the Bekenstein-Hawking
entropy $S_{BH}\roughly>25$ (for $n=6$) in order to ensure that these are close
to having a classical description.  In TeV-gravity/brane world scenarios,
there are two other important approximations.  The first is that the
gravitational field of the brane can be neglected, and is valid for black
holes heavy compared to $M_P$, and the second is to consider
black holes small as compared to the scales of the extra
dimensions.  We may then effectively discuss black holes in $4+n$
dimensional flat space, as studied in \cite{Myers:un}.

The first question in these scenarios concerns the production rate for
black holes.  Two ingredients are needed:  the parton density in a 
proton, which is approximately known, and the
cross-section for two partons to form a black hole, which is not.  This
cross-section may, however, be estimated.  Arguments
along the lines of Thorne's hoop conjecture indicate that a black
hole forms when partons collide at impact parameter $b$ that is less than
the Schwarzschild radius $r_h$ corresponding to $E$.  This would suggest a
parton-parton cross-section of the form
\begin{equation}
\sigma\sim r_h^2(E)\sim E^{1\over {n+1}}\ .\label{crosssec}
\end{equation}
However, until now the high-energy gravitational collision problem has been
little studied.  In 1974, Penrose\cite{Penrose} argued that black holes form in
zero impact parameter collisions, and this work was extended
by D'Eath and Payne\cite{D'Eath:hb,D'Eath:hd,D'Eath:qu}, 
but the problem at non-zero $b$ had not been
systematically treated.  In \cite{Eardley:2002re}, Doug Eardley and I
recently revisited
this problem, and in particular showed that in four dimensions, a
trapped surface forms in collisions with impact parameter
$b\roughly< 1.6E$, very close to the na\"\i ve expectation of
$b\roughly<2E$.  Furthermore, in higher dimensions we reduced the problem
of finding a trapped surface to a higher-dimensional analog of the Plateau
problem, which we expect to have a solution -- work on this continues.

Using the estimate (\ref{crosssec}), one readily finds an impressive
result:  for $M_P\sim 1$ TeV, the LHC will produce black holes with masses
larger than $5 M_P$ at the rate of about one per 
second\cite{Giddings:2001bu,Dimopoulos:2001hw}.  
This would qualify LHC to be called a black
hole factory.

Black holes will then decay leaving spectacular signatures.  The
first stage of their decay is purely classical, and involves the
rather asymmetric initial black hole settling down to a hairless spinning
black
hole, radiating its multipole moments.  We call this stage 
``balding.''  An important open problem is to determine how much
energy is left in the black hole at the end of this stage; rough estimates
based on the size of the initial trapped surface -- which can only grow -- and 
extrapolation of \cite{D'Eath:hb,D'Eath:hd,D'Eath:qu} 
suggest that this energy is around $15-40\%$ of
the initial energy $E$.  We hope that improvement of numerical relativity
or perturbation methods eventually give us a better characterization of
this stage.

Hawking's calculation then becomes relevant.  As in the decay of
four-dimensional black holes, one expects the black hole to first
shed its spin, radiating particles preferentially in the equatorial
plane, in a ``spin-down'' phase.  
Extrapolation of Page's four-dimensional results\cite{Page:df,Page:ki} suggest
a mass loss of perhaps $25\%$ in spin-down.
An important problem is to redo Page's analysis in the higher-dimensional
setting.

Spin-down leaves a Schwarzschild black hole which continues to evaporate
through the ``Schwarzschild phase.''  The instantaneous energy
distribution is thermal, and may be integrated to find an overall
spectrum.  This phase should represent perhaps $75\%$ of the black hole
decay energy.

When the black hole reaches the Planck size, we confront the
profound mystery we began with:  what effects govern the final decay, 
what do they tell us about the nature of
quantum gravity, and what happens to information?  
Exploration of this ``Planck phase'' is beyond present theoretical
technology, which makes the prospect of experimental results all the more
tantalizing.

Products of these stages should stand out in 
accelerators\cite{Giddings:2001bu,Dimopoulos:2001hw}.  In particular, a black
hole should produce of order $S_{BH}$ energetic primary particles -- leptons,
quarks, gluons, {\it etc.} -- in its Schwarzschild phase.  
These will be radiated
roughly isotropically, with characteristic spectra and species ratios,
as predicted by Hawking's calculations and numerical extensions.
These events should not be masked by backgrounds from any known extrapolation
of the standard model -- they are very unique.  

This scenario has other interesting consequences.  First, we know 
that cosmic rays hit our atmosphere with
center-of-mass energies exceeding energies accessible at LHC:  if TeV-scale
gravity is correct, black hole events have peppered our upper
atmosphere throughout earth's history.  If ultrahigh-energy
neutrino cosmic ray fluxes are sufficiently strong, these events may even be
observable in the next round of cosmic ray 
observatories\cite{Feng:2001ib,Anchordoqui:2001ei,Emparan:2001kf,Giddings:2001ih,
Ringwald:2002vk}.

It also appears likely that, through the AdS/CFT correspondence,
relativists might tell particle theorists something about QCD.  I have
recently argued\cite{Giddings:2002cd} 
that black hole formation or other strong gravity effects
in anti-de Sitter space are dual to the physics that saturates the
Froissart bound for hadron cross-sections
\begin{equation}
\sigma \sim \ln^2 E\ .
\end{equation}
Important questions remain regarding the structure and stability of such
black holes.

Finally, consequences for the future of high energy physics are
profound.  Humanity has pursued a longstanding quest to understand physics at
shorter and shorter distances.  In quantum gravity we don't
know that distances shorter than the Planck scale exist, but this is a
question that should be addressed experimentally.  However, once we
start making black holes this appears impossible.  Any attempt at
collisions that can probe shorter distances will be cloaked inside
an event horizon, and all that will be seen in our detectors is
the products of the black hole decay -- there is apparently no way to
directly observe the short-distance physics taking place inside the black
hole.  Black hole production represents the end of short-distance physics.
However, it doesn't necessarily spell a dismal future for high-energy
experiments.  As we produce bigger black holes, they reach off
the brane that is our world and offer us a way to explore the geometry and
other features of the extra dimensions.  High energy physics can become the
study of the geography of extra dimensions.

\bigskip
\bigskip
\centerline{\bf Acknowledgments}
\bigskip

I'd like to thank my collaborators D. Eardley, E. Katz, and S. Thomas for
the opportunity to explore these fascinating ideas together.  This work was
supported in part by the Department of Energy under Contract
DE-FG-03-91ER40618, and by 
David and Lucile Packard Foundation Fellowship
2000-13856.

\bibliography{paper}

\begin{thebibliography}{99}
\baselineskip=14pt

\bibitem{Hawking:sw}
S.~W.~Hawking,
``Particle Creation By Black Holes,''
Commun.\ Math.\ Phys.\  {\bf 43}, 199 (1975).
%%CITATION = CMPHA,43,199;%%

\bibitem{Hawking:ra}
S.~W.~Hawking,
``Breakdown Of Predictability In Gravitational Collapse,''
Phys.\ Rev.\ D {\bf 14}, 2460 (1976).
%%CITATION = PHRVA,D14,2460;%%

\bibitem{Giddings:1995gd}
S.~B.~Giddings,
``The Black hole information paradox,''
arXiv:hep-th/9508151.
%%CITATION = HEP-TH 9508151;%%

\bibitem{Strominger:1994tn}
A.~Strominger,
``Les Houches lectures on black holes,''
arXiv:hep-th/9501071.
%%CITATION = HEP-TH 9501071;%%

\bibitem{Giddings:1994pj}
S.~B.~Giddings,
``Quantum mechanics of black holes,''
arXiv:hep-th/9412138.
%%CITATION = HEP-TH 9412138;%%

\bibitem{tHooft:ij}
G.~'t Hooft,
``Black Holes, Hawking Radiation, And The Information Paradox,''
Nucl.\ Phys.\ Proc.\ Suppl.\  {\bf 43}, 1 (1995).
%%CITATION = NUPHZ,43,1;%%

\bibitem{Giddings:2001bu}
S.~B.~Giddings and S.~Thomas,
``High energy colliders as black hole factories: The end of short  distance physics,''
Phys.\ Rev.\ D {\bf 65}, 056010 (2002)
[arXiv:hep-ph/0106219].
%%CITATION = HEP-PH 0106219;%%

\bibitem{Dimopoulos:2001hw}
S.~Dimopoulos and G.~Landsberg,
``Black holes at the LHC,''
Phys.\ Rev.\ Lett.\  {\bf 87}, 161602 (2001)
[arXiv:hep-ph/0106295].
%%CITATION = HEP-PH 0106295;%%

\bibitem{Myers:un}
R.~C.~Myers and M.~J.~Perry,
``Black Holes In Higher Dimensional Space-Times,''
Annals Phys.\  {\bf 172}, 304 (1986).
%%CITATION = APNYA,172,304;%%

\bibitem{Penrose}
R.Penrose
{\sl unpublished} (1974).

\bibitem{D'Eath:hb}
P.~D.~D'Eath and P.~N.~Payne,
``Gravitational Radiation In High Speed 
Black Hole Collisions. 1. Perturbation Treatment Of 
The Axisymmetric Speed Of Light Collision,''
Phys.\ Rev.\ D {\bf 46}, 658 (1992).
%%CITATION = PHRVA,D46,658;%%

\bibitem{D'Eath:hd}
P.~D.~D'Eath and P.~N.~Payne,
``Gravitational Radiation In High Speed Black Hole Collisions. 
2. Reduction To Two Independent Variables And Calculation Of The Second Order
News Function,''
Phys.\ Rev.\ D {\bf 46}, 675 (1992).
%%CITATION = PHRVA,D46,675;%%

\bibitem{D'Eath:qu}
P.~D.~D'Eath and P.~N.~Payne,
``Gravitational Radiation In High Speed Black Hole Collisions. 3. 
Results And Conclusions,''
Phys.\ Rev.\ D {\bf 46}, 694 (1992).
%%CITATION = PHRVA,D46,694;%%

%\cite{Eardley:2002re}
\bibitem{Eardley:2002re}
D.~M.~Eardley and S.~B.~Giddings,
``Classical black hole production in high-energy collisions,''
arXiv:gr-qc/0201034.
%%CITATION = GR-QC 0201034;%%

\bibitem{Page:df}
D.~N.~Page,
``Particle Emission Rates From A Black Hole: 
Massless Particles From An Uncharged, Nonrotating Hole,''
Phys.\ Rev.\ D {\bf 13}, 198 (1976).
%%CITATION = PHRVA,D13,198;%%

\bibitem{Page:ki}
D.~N.~Page,
``Particle Emission Rates From A Black Hole. II. 
Massless Particles From A Rotating Hole,''
Phys.\ Rev.\ D {\bf 14}, 3260 (1976).
%%CITATION = PHRVA,D14,3260;%%

\bibitem{Feng:2001ib}
J.~L.~Feng and A.~D.~Shapere,
``Black hole production by cosmic rays,''
arXiv:hep-ph/0109106.
%%CITATION = HEP-PH 0109106;%%


\bibitem{Anchordoqui:2001ei}
L.~Anchordoqui and H.~Goldberg,
``Experimental signature for black hole production in neutrino air  showers,''
arXiv:hep-ph/0109242.
%%CITATION = HEP-PH 0109242;%%

\bibitem{Emparan:2001kf}
R.~Emparan, M.~Masip and R.~Rattazzi,
``Cosmic rays as probes of large extra dimensions and TeV gravity,''
arXiv:hep-ph/0109287.
%%CITATION = HEP-PH 0109287;%%

\bibitem{Giddings:2001ih}
S.~B.~Giddings,
``Black hole production in TeV-scale gravity, and the future of high  energy physics,''
in {\it Proc. of the APS/DPF/DPB Summer Study on the Future of Particle Physics (Snowmass 2001) } ed. R.~Davidson and C.~Quigg,
arXiv:hep-ph/0110127.
%%CITATION = HEP-PH 0110127;%%


\bibitem{Ringwald:2002vk}
A.~Ringwald and H.~Tu,
``Collider versus cosmic ray sensitivity to black hole production,''
Phys.\ Lett.\ B {\bf 525}, 135 (2002)
[arXiv:hep-ph/0111042].
%%CITATION = HEP-PH 0111042;%%

\bibitem{Giddings:2002cd}
S.~B.~Giddings,
``High energy QCD scattering, the shape of gravity on an IR brane, and  the Froissart bound,''
arXiv:hep-th/0203004.
%%CITATION = HEP-TH 0203004;%%


\end{thebibliography}

\end{document}